\documentclass{jetplwin}
\usepackage{amsmath}
\twocolumn \rus

\title{Anisotropic 2D Larkin-Imry-Ma state in polar distorted ABM phase of $^3$He in ''nematically ordered'' aerogel}

\rtitle{Anisotropic 2D Larkin-Imry-Ma state in polar distorted ABM
phase of $^3$He in ''nematically ordered'' aerogel}

\sodtitle{Anisotropic 2D Larkin-Imry-Ma state polar distorted ABM
phase of $^3$He in ''nematically ordered'' aerogel}

\author{R.\,Sh.\,Askhadullin$^+$, V.\,V.\,Dmitriev\thanks{e-mail: dmitriev@kapitza.ras.ru},
P.\,N.\,Martynov$^+$, A.\,A.\,Osipov$^+$, A.\,A.\,Senin,
A.\,N.\,Yudin}

\rauthor{R.\,Sh.\,Askhadullin$^+$, V.\,V.\,Dmitriev,
P.\,N.\,Martynov$^+$, A.\,A.\,Osipov$^+$, A.\,A.\,Senin,
A.\,N.\,Yudin}

\sodauthor{R.\,Sh.\,Askhadullin$^+$, V.\,V.\,Dmitriev,
P.\,N.\,Martynov$^+$, A.\,A.\,Osipov$^+$, A.\,A.\,Senin,
A.\,N.\,Yudin}

\address{P.\,L. Kapitza Institute for Physical Problems of RAS,
2 Kosygina str., 119334 Moscow, Russia\\~\\
$^+$ A.\,I.\,Leypunsky Institute for Physics and Power
Engineering, Obninsk, Kaluga region, Russia}

\dates{20 October 2014}{*}

\abstract{We present results of experiments in superfluid phases
of $^3$He confined in aerogel which strands are nearly parallel to
one another. High temperature superfluid phases of $^3$He in this
aerogel (ESP1 and ESP2) are chiral phases and have polar distorted
ABM order parameter which orbital part forms 2D Larkin-Imry-Ma
state. We demonstrate that this state can be anisotropic if the
aerogel is squeezed in direction transverse to the strands. Values
of this anisotropy in ESP1 and ESP2 phases are different, what
leads to different NMR properties.}

\PACS{67.57.Pq, 67.57.Lm}

\begin{document}
\maketitle
\section{Introduction}

A so-called ``nematically ordered'' (N-) aerogel differs from
standard silica aerogels by a high value of a global anisotropy.
This aerogel consists of Al$_2$O$_3\cdot$H$_2$O strands which are
nearly parallel to one another \cite{Askh}, i.e. it may be
considered as aerogel with infinite stretching anisotropy.
Investigations of superfluid $^3$He confined in N-aerogel are
especially interesting because according to a theory \cite{Aoyama}
such a strong anisotropy may make a superfluid polar phase more
favorable than Anderson-Brinkman-Morel (ABM) phase which
corresponds to A phase of bulk $^3$He and to A-like phase of
$^3$He in isotropic or weakly anisotropic silica aerogels
\cite{Kun, Dmit2010, Halp1}. A superfluid phase diagram of $^3$He
in N-aerogel is different from the case of $^3$He in silica
aerogel with similar porosity \cite{we12}. The superfluid
transition temperature ($T_{ca}$) is slightly (by 3-6\%)
suppressed in comparison with the transition temperature ($T_c$)
of bulk $^3$He. Depending on prehistory, pressure and temperature,
three superfluid phases are observed: two Equal Spin Pairing
phases (ESP1 or ESP2) and Low Temperature phase (LTP). The ESP1
phase appears on cooling from the normal state. On further cooling
the first order transition into the LTP takes place. Due to
inhomogeneities of the aerogel, this transition occurs in a wide
temperature range ($\sim0.05\,T_c$). On warming from the LTP the
back transition into the ESP phase is observed. At high pressures
($P\geq 10$\,bar) the NMR frequency shift in this phase, called
ESP2 phase, is greater than in the ESP1 phase at the same
conditions.

The LTP has a polar distorted Balian-Werthamer (BW) order
parameter \cite{we14}. As for ESP phases, their NMR properties
point out that they both have ABM order parameter with a strong
polar distortion \cite{we12}. This distortion is larger at low
pressures and at higher temperatures. It was also found that the
order parameter orbital vector ${\bf l}$ of the distorted ABM
phase in N-aerogel is in a spatially inhomogeneous Larkin-Imry-Ma
(LIM) state similar to that predicted in \cite{Vol} and observed
in A-like phase of $^3$He in silica aerogel \cite{Dmit2010,
Halp2}. In N-aerogel we get the two-dimensional LIM state because
the aligned strands orient ${\bf l}$ normal to their axis.

In this paper we present results of nuclear magnetic resonance
(NMR) studies of liquid $^3$He confined in N-aerogel which was
slightly squeezed in direction transverse to the strands. In
particular, these experiments allow us to explain the difference
between properties of ESP1 and ESP2 phases.

\section{Theory}

Transverse NMR frequency shift can be found from the following
equation \cite{Fom0}:
\begin{equation}
\Delta\omega = -\frac{g}{\chi H}\frac{\partial
\bar{U}_D}{\partial\cos\beta} , \label{equ0}
\end{equation}
where $g$ is the gyromagnetic ratio, $\chi$ - the spin
susceptibility, $H$ - the external magnetic field, $\beta$ - the
tipping angle of the magnetization and $\bar{U}_D$ - the density
of the dipole energy, averaged over a fast spin precession. For
the LIM state the dipole energy should also be averaged over the
space (see e.g. \cite{Dmit2010, BK}). The order parameter of the
ABM phase with polar distortion is:
\begin{equation}
A_{jk} =\Delta_0 e^{i\phi}d_j\left(am_k+ibn_k\right), \label{eq1}
\end{equation}
where $\Delta_0$ is the gap parameter, ${\bf d}$ is the unit spin
vector, ${\bf m}$ and ${\bf n}$ are mutually orthogonal unit
vectors in the orbital space and $a^2+b^2=1$. For the ABM phase
$a=b$, for polar distorted ABM phase $a^2>b^2$ and for polar phase
$a=1, b=0$. Similarly to pure ABM phase, the distorted ABM phase
is a chiral phase and we can introduce the orbital vector ${\bf
l}={\bf m}\times{\bf n}$ which orientation defines two Weyl points
in the momentum space: the energy gap of this phase equals 0 along
${\bf l}$ and equals $\sqrt{2}a\Delta_0$ and $\sqrt{2}b\Delta_0$
along $\bf m$ and $\bf n$. Note that the polar phase is not chiral
and its gap has line of zeroes in the plane normal to {\bf m}. The
dipole energy density for the order parameter \eqref{eq1} is:
\begin{equation}
U_{D} =\frac{6}{5}g_D\left(a^2({\bf dm})^2+b^2({\bf dn})^2\right),
\label{equ2}
\end{equation}
where $g_D=g_D(T)$ is the dipole constant. In weak coupling limit
$g_D$ can be expressed in terms of the Leggett frequency of the
pure ABM phase $\Omega_A$ \cite{VW}:
\begin{equation}
g_{D}
=\frac{2}{3-4a^2b^2}~g_D^{A}=\frac{2}{3-4a^2b^2}\Big(\frac{5}{6}\frac{\chi}{g^2}\Omega^2_A\Big),
\label{equ2a}
\end{equation}
where $g_D^{A}$ is the dipole constant of the ABM phase. Strong
coupling corrections to \eqref{equ2a} do not exceed $\pm$5\%
\cite{Min}, therefore we do not consider them below.

Following \cite{BK,BK2}, we use two coordinate frames: an orbital
frame ($\hat \xi, \hat \eta, \hat \zeta$) bound to the aerogel
sample and a spin frame ($\hat x, \hat y, \hat z$). We choose
${\bf H}=H\hat z$ and fix $\hat\zeta$-axis along aerogel strands.
Then strands of N-aerogel orient $\bf{m}\parallel\hat\zeta$ and
${\bf l}\perp\hat \zeta$ \cite{Aoyama}. In the isotropic 2D LIM
state vectors ${\bf l}$ and ${\bf n}$ are randomly distributed in
$\hat\xi-\hat\eta$ plane and $\left<l_\xi^2\right> =
\left<l_\eta^2\right> = \left<n_\xi^2\right> =
\left<n_\eta^2\right> =\frac{1}{2}$, where angle brackets mean the
space averaging. We introduce the angle $\lambda=\lambda({\bf r})$
which defines the orientation of specific $\bf l$ and the
corresponding $\bf n$: $l_\xi = - n_\eta = \cos\lambda$ and
$l_\eta = n_\xi = \sin\lambda$. For uniaxially anisotropic in
$\hat{\xi}$-$\hat{\eta}$ plane 2D LIM state we fix the
$\hat{\xi}$-axis along the direction corresponding to the maximum
value of $\left<l_\xi^2\right>$. Consequently
$1>\left<l_\xi^2\right>
>\frac{1}{2} > \left<l_\eta^2\right>$ and we assume that
the distribution of $l_\xi =l_\xi(\lambda)$ is symmetric.

\begin{figure}[t]
\centerline{\includegraphics[width=2.5in]{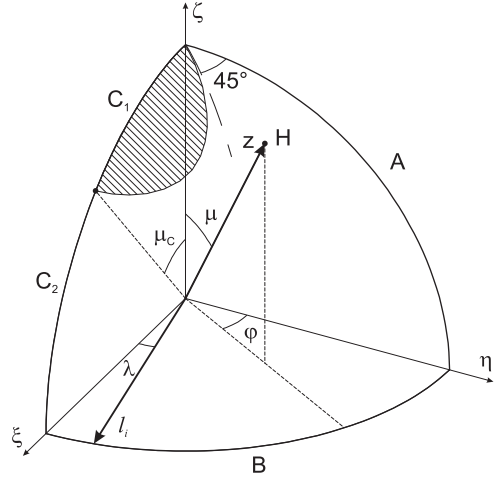}} \caption{Fig.1.
Orientation of $\bf H$ with respect to N-aerogel axes.} \label{f1}
\end{figure}

An orientation of $\bf H$ with respect to the aerogel
(Fig.\ref{f1}) is described by angles of rotation of the orbital
frame: $\mu$ (rotation around $\hat \xi$) and $\varphi$ (rotation
around $\hat \zeta$). Then we get:
\begin{equation}
\begin{array}{lcl}
m_x=0,~~m_y=-\sin\mu,~~m_z=\cos\mu,\\
n_x=\sin(\varphi+\lambda),~~n_y=-\cos\mu\cos(\varphi+\lambda),\\
n_z=-\sin\mu\cos(\varphi+\lambda). \label{equ2b}
\end{array}
\end{equation}

Motions of $\bf d$ in the spin frame are described by Euler angles
($\alpha, \beta, \gamma$), where $\alpha$ corresponds to the phase
of spin precession and $\beta$ is the tipping angle. After an
averaging over the fast spin precession we obtain:
\begin{equation}
\begin{array}{lcl}
\bar{d}_x^2
=\frac{1}{4}\left<\cos^2\Phi\right>(1+\cos\beta)^2+\frac{1}{8}(1-\cos\beta)^2,\\
\\
\bar{d}_y^2=
\frac{1}{4}\left<\sin^2\Phi\right>(1+\cos\beta)^2+\frac{1}{8}(1-\cos\beta)^2,\\
\\
\bar{d}_z^2=\frac{1}{2}\sin^2\beta,~~~\overline{d_xd_z}=\overline{d_yd_z}=0,\\
\\
\overline{d_xd_y}
=-\frac{1}{8}\left<\sin2\Phi\right>(1+\cos\beta)^2,\\
\label{equ3}
\end{array}
\end{equation}
where $\Phi=\alpha+\gamma$ is a slow variable. Then the dipole
energy \eqref{equ2} averaged over the space is:
\begin{eqnarray}
\bar{U}_{D}
=\frac{6}{5}g_D\Big[a^2(\bar{d}_y^2m_y^2+\bar{d}_z^2m_z^2)+~~~~~~~~~~~
\nonumber
\\
+~b^2(\bar{d}_x^2\left<n_x^2\right>+\bar{d}_y^2\left<n_y^2\right>+\bar{d}_z^2\left<n_z^2\right>+
2\overline{d_xd_y}\left<n_xn_y\right>)\Big], \label{equ3a}
\end{eqnarray}
where
$\left<n_x^2\right>=\sin^2\varphi\left<\cos^2\lambda\right>+\cos^2\varphi\left<\sin^2\lambda\right>$,
$\left<n_y^2\right>=\cos^2\mu(\cos^2\varphi\left<\cos^2\lambda\right>+\sin^2\varphi\left<\sin^2\lambda\right>$),
$\left<n_z^2\right>=\sin^2\mu(\cos^2\varphi\left<\cos^2\lambda\right>+\sin^2\varphi\left<\sin^2\lambda\right>$)
and
$\left<n_xn_y\right>=(2\left<\sin^2\lambda\right>-1)\cos\mu\sin\varphi\cos\varphi$.
The angle $\Phi$ may be spatially homogeneous (the spin nematic
state, SN) or random (the spin glass state, SG) \cite{Dmit2010}.
The SN state is more favorable and corresponds to the homogeneous
spatial distribution of ${\bf d}$, but the SG state may be created
e.g. in pulse NMR experiments after an application of large
tipping pulses. In the isotropic SG state
$\left<\sin^2\Phi\right>=\left<\cos^2\Phi\right>=1/2$ and
$\left<\sin2\Phi\right>=0$ while in the SN state $\Phi$ is
determined by minimization of \eqref{equ3a}. The result of the
minimization is shown in Fig.1 where the shaded area corresponds
to orientations of $\bf H$ with $\sin^2\Phi=1$ while for other
orientations the minimum of \eqref{equ3a} corresponds to
$\sin^2\Phi=0$. The border of the shaded area satisfies to the
following condition:
\begin{eqnarray}
b^2\big((\left<l_\xi^2\right>\cos^2\varphi+\left<l_\eta^2\right>\sin^2\varphi)\cos^2\mu-~~~~~~
\nonumber \\
-\left<l_\xi^2\right>\sin^2\varphi-\left<l_\eta^2\right>\cos^2\varphi\big)+a^2\sin^2\mu=0.
\label{equ4}
\end{eqnarray}
In particular, if $\varphi=90^\circ$ then $\sin^2\Phi=1$ for
$\mu<\mu_c$ and $\sin^2\Phi=0$ for $\mu>\mu_c$, where
\begin{equation}
\sin\mu_c
=\frac{b^2(1-2\left<l_\eta^2\right>)}{1-b^2-b^2\left<l_\eta^2\right>}~.
\label{equ5}
\end{equation}
The critical angle $\mu_c$ corresponds to an orientational
transition: in the equilibrium SN state ${\bf d}\perp\hat\eta$ for
$\mu<\mu_c$, while ${\bf d}\parallel\hat\eta$ for $\mu>\mu_c$.

\begin{figure}[t]
\centerline{\includegraphics[width=2.8in]{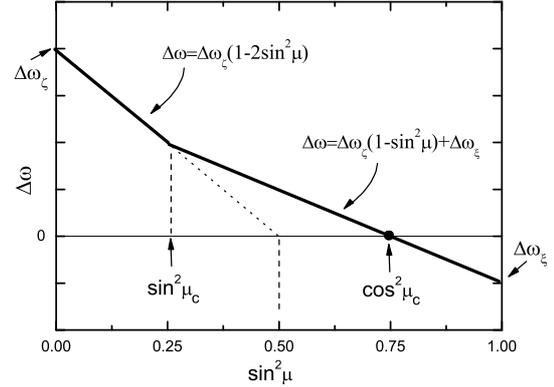}} \caption{Fig.2.
CW NMR frequency shift versus $\mu$ for $\varphi=90^\circ$ as
follows from \eqref{equ6}.} \label{f2}
\end{figure}

The NMR frequency shift from the Larmor value can be obtained from
\eqref{equ0} and \eqref{equ3a}:
\begin{eqnarray}
\Delta\omega=\frac{1}{4}K
\Big[\big(a^2m^2_y-b^2\left<n^2_x\right>+b^2\left<n^2_y\right>\big)\times~~~~~~~~
\nonumber \\
\times\big(1-2\sin^2\Phi(1+\cos\beta)\big)+ ~~~~~~~~~~
\nonumber \\
~~~+~
\big(4-5a^2m^2_y-b^2(7\left<n^2_x\right>+5\left<n^2_y\right>)\big)\cos\beta\Big],
\label{equ5a}
\end{eqnarray}
where $$K=\frac{2}{3-4a^2b^2}\frac{\Omega^2_A}{\omega}$$ and
$\omega=gH$. Let consider 4 cases: $\varphi=0$, $0<\mu<90^\circ$
(the case A); $\mu=90^\circ$, $0<\varphi<90^\circ$ (B);
$\varphi=90^\circ$, $0<\mu<\mu_c$ (C$_1$); $\varphi=90^\circ$,
$\mu_c<\mu<90^\circ$ (C$_2$). In Fig.\ref{f1} these orientations
of $\bf H$ correspond to arcs marked $A,~B,~C_1$ and $C_2$. Then
for the case of continuous wave (CW) NMR ($\cos\beta\approx 1$) we
get:
\begin{equation}
\begin{array}{lcl}
A:~~\Delta\omega=K(D\sin^2\mu+E\cos^2\mu),\\
B:~~\Delta\omega=KD(1-2\sin^2\varphi),\\
C_1:~~\Delta\omega=KE\cos2\mu,\\
C_2:~~\Delta\omega=K(E\cos^2\mu-D), \label{equ6}
\end{array}
\end{equation}
where $D=b^2(1-2\left<l_\eta^2\right>)\geq0$ and
$E=1-b^2-b^2\left<l_\eta^2\right>>0$. The dependence of
$\Delta\omega$ on $\mu$ for $\varphi=90^\circ$ is shown in
Fig.\ref{f2}. This dependence is fully determined by 2 values of
the frequency shift: $\Delta\omega_{\xi}=-KD$ (${\bf
H}\parallel\hat\xi$) and $\Delta\omega_\zeta=KE$ (${\bf
H}\parallel\hat\zeta$) so that
$\sin^2\mu_c=-\Delta\omega_\xi/\Delta\omega_\zeta$.

In the isotropic 2D LIM state $\left<l_\eta^2\right>=1/2$ (i.e.
$D=0$) and for $\mu=90^\circ$ (the case B) $\Delta\omega=0$ in
agreement with \cite{we12}. If the 2D LIM state is anisotropic and
$\left<l_\eta^2\right><1/2$, then for $\mu=90^\circ$ the shift
equals 0 for $\varphi=45^\circ$. For other values of $\varphi$ the
shift is 0 only for pure polar phase ($b=0$). In pure ABM phase or
in the ABM phase with polar distortion the shift is positive (if
$\varphi<45^\circ$) or negative (if $\varphi>45^\circ$).

\section{Experimental setup}

The experimental chamber used in the present work is similar to
the chamber described in \cite{we12}. The chamber has two cells
with N-aerogel samples. The samples (named below as 1 and 2) have
a form of a cuboid with characteristic sizes of 4\,mm. Initially
the samples had an overall density $\sim$30\,mg/cm$^2$ (sample 1)
and $\sim$8\,mg/cm$^2$ (sample 2), but were squeezed by $\sim$10\%
and $\sim$5\% correspondingly along the direction transverse to
the aerogel strands. In order to match Fig.\ref{f1} we choose the
direction of the squeezing as $\eta$-axis, because (see next
section) in the anisotropic 2D LIM state of the distorted ABM
phase the maximum of $\left<l_\xi^2\right>$ corresponds to this
direction of the squeezing.

We were able to rotate $\bf H$ by any angle $\mu$ in $\hat
\zeta$-$\hat\eta$ plane (for the sample 1) or in $\hat
\zeta$-$\hat\xi$ plane (for the sample 2). Additional gradient
coils were used to compensate the external magnetic field
inhomogeneity. Experiments were performed in magnetic fields from
104\,Oe up to 425\,Oe (NMR frequencies were from 340\,kHz up to
1.38\,MHz) and at pressures from s.v.p. up to 29.3\,bar. The
necessary temperatures were obtained by a nuclear demagnetization
cryostat and were measured by a quartz tuning fork, calibrated by
Leggett frequency measurements in bulk $^3$He-B. In order to avoid
a paramagnetic signal from surface solid $^3$He, the samples were
preplated by $\sim$2.5 atomic monolayers of $^4$He.

A superfluid phase diagram of $^3$He in the sample 1 was found to
be almost the same as the diagram presented in \cite{we12}. For
the sample 2 the diagram is slightly different (the superfluid
transition temperatures are by 2-3\% higher).

\section{Experiments with the sample 1}
Temperature dependencies of CW NMR frequency shifts in the ESP1
phase ($\Delta\omega_\zeta$ and $\Delta\omega_\eta$) for ${\bf
H}\parallel\hat\zeta$ and for ${\bf H}\parallel\hat\eta$ are shown
in Fig.\ref{f3}. The superfluid transition temperature of $^3$He
in this sample at the given pressure (14.2\,bar) is
$\sim$0.965\,$T_c$ as it can be seen from appearance of the NMR
shift for ${\bf H}\parallel\hat\zeta$. As follows from
\eqref{equ6} $\Delta\omega_\zeta=KE$ and $\Delta\omega_\eta=KD$.
In the experiment, we obtain $\Delta\omega_\eta=0$ down to
$\sim0.93\,T_c$ but on further cooling the positive shift appears.
It means that for $T<0.93\,T_c$ both $b^2$ and
$(1-2\left<l_\eta^2\right>)$ are nonzero and positive, i.e. we get
the distorted ABM phase and the squeezing of the sample along
$\hat\eta$ results in preferable orientation of vectors $\bf l$
along the $\hat\xi$-axis.

If value of $\Omega_A$ is known, then we can find $b^2$ and
$\left<l_\eta^2\right>$ from the measured values of
$\Delta\omega_\zeta$ and $\Delta\omega_\eta$. The problem is that
$\Omega_A\propto \Delta_0$ is known only for the bulk $^3$He
(below we denote this value by $\Omega_{A0}$). The value of
$\Omega_A$ in N-aerogel should be smaller due to the suppression
of $T_c$ and corresponding decrease of the gap. It is known that
in silica aerogels the gap suppression is larger than the
suppression of $T_c$ in agreement with the ``inhomogeneous
isotropic scattering model'' \cite{IISM}. For example, for
$T_{ca}=0.965\,T_c$ the gap and $\Omega_A$ is suppressed by
$\sim$9\% \cite{HJ}. Another model called the ``homogeneous
isotropic scattering model'' \cite{HISM} predicts that the gap
suppression is proportional to $T_{ca}/T_c$. Both these models,
however, can not be directly applied to $^3$He in N-aerogel due to
its strong anisotropy. Therefore we can only assume that the
suppression of $\Omega_A$ in N-aerogel is proportional to
$T_{ca}/T_c$ or larger, i.e.:

\begin{figure}[t]
\centerline{\includegraphics[width=2.8in]{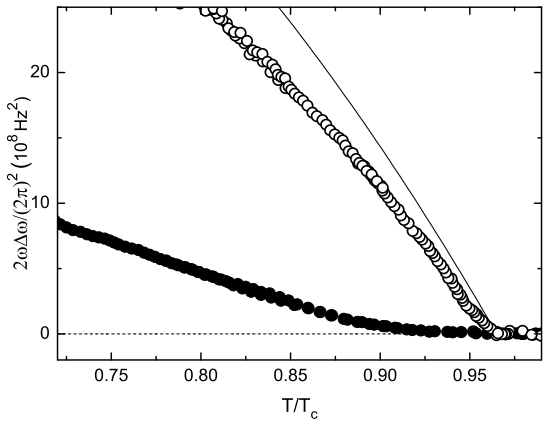}} \caption{Fig.3.
CW NMR frequency shift versus temperature in sample 1.
({\Large$\circ$}) -- ${\bf H}\parallel\hat\zeta$;
({\Large$\bullet$}) -- ${\bf H}\parallel\hat\eta$. Solid line
shows $\Omega_A^2$ rescaled from $\Omega_{A0}^2$ in accordance
with \eqref{equ12} for $k=1$. $P$ = 14.2\,bar,
$T_{ca}\approx0.965\,T_c$.} \label{f3}
\end{figure}

\begin{equation}
\Omega_A({T}/{T_{ca}})=k\frac{T_{ca}}{T_c} \Omega_{A0}({T}/{T_c}),
\label{equ12}
\end{equation}
where $k\leq 1$. In Fig.\ref{f4} we present values of $b^2$ and
$\left<l_\eta^2\right>$ calculated from the data shown in
Fig.\ref{f3} in the assumption that $k=1$ or $k=0.9$. It can be
seen that $\left<l_\eta^2\right>$ grows on warming and tends to
1/2 for both values of $k$, while $b^2$ decreases but can not be
extrapolated to 0 for $T<T_{ca}$. Thus we conclude that the
anisotropy of the 2D LIM state decreases on warming and
$\left<l_\eta^2\right>$ becomes equal (or close) to 1/2 at
$T>0.93\,T_{ca}$ resulting in $\Delta\omega_\eta=0$. The polar
distortion grows on warming but it is unlikely that we get pure
polar phase in a reasonably large temperature range near $T_{ca}$
for these values of $k$. At lower pressures ($P\leq9$\,bar) we
have obtained similar dependencies as shown in Figs.\ref{f2} and
\ref{f3}, but if $k\leq0.9$ the value of $b^2$ can be extrapolated
to 0 at $T<T_{ca}$, so the existence of the pure polar phase near
$T_{ca}$ can not be excluded, but only if $k\leq0.9$.

\begin{figure}[t]
\centerline{\includegraphics[width=2.7in]{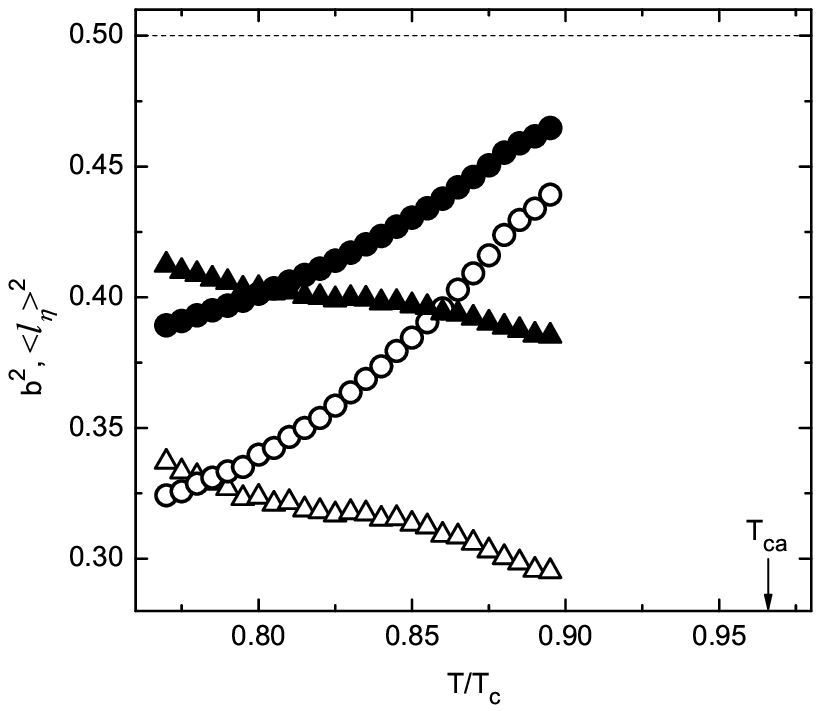}} \caption{Fig.4.
$\left<l_\eta^2\right>$ (circles) and $b^2$ (triangles) calculated
from data in Fig.\ref{f3} for $k=1$ ({\Large$\bullet$},
{\large{$\blacktriangle$}}) and for $k=0.9$ ({\Large$\circ$},
{\small$\triangle$}).} \label{f4}
\end{figure}

\section{Identification of the ESP2 phase (Experiments with the sample 2)}
\begin{figure}[t]
\centerline{\includegraphics[width=2.7in]{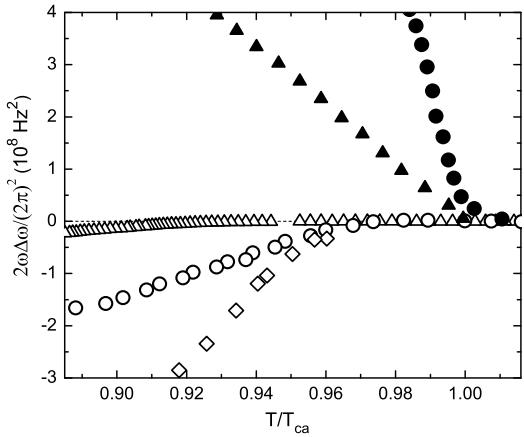}} \caption{Fig.5.
CW NMR frequency shift versus temperature for sample 2. Open
symbols -- ${\bf H}\parallel\hat\xi$; filled symbols -- ${\bf
H}\parallel\hat\zeta$. $P=12.3$\,bar ($T_{ca}\approx 0.98\,T_c$):
({\Large$\bullet$}, {\Large$\circ$}) -- ESP1 phase;
({\Large{$\diamond$}}) -- ESP2 phase.~~~ $P=2.9$\,bar
($T_{ca}\approx 0.95\,T_c$): ({\large{$\blacktriangle$}},
{\small{$\triangle$}}) -- ESP1 phase. } \label{f5}
\end{figure}

The sample 2 was oriented so that $\bf H$ could be rotated in
$\hat \zeta$-$\hat \xi$ plane. Correspondingly, at low
temperatures the anisotropy of the 2D LIM state should result in a
negative CW NMR frequency shift for the transverse orientation of
the field (${\bf H}\parallel\hat\xi$) as follows from $C_2$\ in
\eqref{equ6} for $\mu=90^\circ$ and $D>0$. Examples of temperature
dependencies of the shift in ESP phases for both transverse and
parallel orientations of ${\bf H}$ are shown in Fig.\ref{f5}. As
it was expected, at low temperatures the shift is negative in both
ESP phases for ${\bf H}\parallel\hat\xi$. In this case the
absolute value of the shift in the ESP2 phase is larger than in
the ESP1 phase.

The difference in the NMR shift in ESP1 and ESP2 phases may be
explained either by different values of the polar distortion (i.e.
of $b^2$) or by different values of the anisotropy of the 2D LIM
state (i.e. of $\left<l_\eta^2\right>$). In both cases the
dependence of $\Delta\omega=\Delta\omega(\mu)$ for the sample 2
should correspond to the dependence shown in Fig.\ref{f2}.

\begin{figure}[t]
\centerline{\includegraphics[width=2.8in]{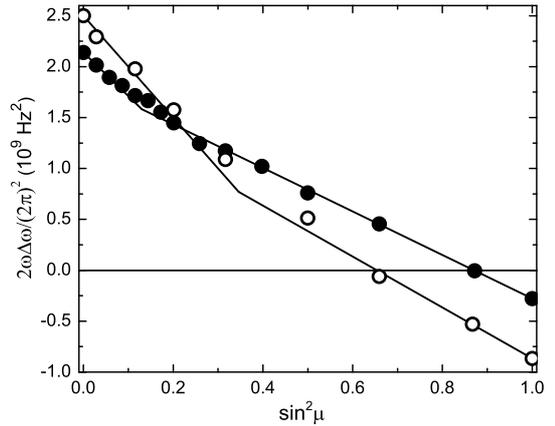}} \caption{Fig.6.
CW NMR frequency shift versus $\mu$ for sample 2. Solid lines --
theoretical dependencies $C_1$ and $C_2$ in \eqref{equ6}.
({\Large$\bullet$}) -- ESP1 phase; ({\Large$\circ$}) -- ESP2
phase. $P=12.3$\,bar, $T\approx 0.85\,T_{ca}$, $H=$117\,Oe.}
\label{f6}
\end{figure}

As follows from \eqref{equ5} $\mu_c>0$ only if $\Delta
\omega_{\xi}\neq 0$. Therefore we have chosen the temperature
$T\approx 0.85\,T_{ca}$ where the absolute value of $\Delta
\omega_\xi$ is large enough, but the smeared transition into the
LTP just starts. In order to get the ESP2 phase at this
temperature the sample was warmed up above the point of full
transition to the ESP2 phase ($\sim0.93\,T_{ca}$) and then was
cooled down. The obtained dependencies of the CW NMR shift on
$\mu$ are shown in Fig.\ref{f6} where solid lines are drawn using
only the corresponding values of $\Delta\omega_\xi$ and
$\Delta\omega_\zeta$. It can be seen that the data are well
described by the theory. Further analysis shows that the
difference between the ESP phases can not be attributed to the
difference of magnitudes of the polar distortion, but can be
explained in assumption that the anisotropy of the 2D LIM state in
these phases is different. The data in Fig.\ref{f6} allow to
calculate $b^2$ and $\left<l_\eta^2\right>$ for a given $k$ in
Eq.\eqref{equ12}. If $k=1$ then $b^2=0.42$ and
$\left<l_\eta^2\right>=0.43$ for the ESP1 phase, and $b^2=0.43$
and $\left<l_\eta^2\right>=0.33$ for ESP2 phase. For $k=0.9$ we
get $b^2=0.35$ and $\left<l_\eta^2\right>=0.39$ for the ESP1
phase, and $b^2=0.36$ and $\left<l_\eta^2\right>=0.24$ for ESP2
phase. Note that for both values of $k$ we get nearly equal values
of $b^2$ in both ESP states, while the anisotropy of the 2D LIM
state in the ESP2 phase is always greater than in the ESP1 phase.
This difference in the anisotropy may be due to the ESP2 phase is
formed on warming from the LTP, which order parameter corresponds
to a spatially homogeneous polar distorted BW phase. It is the
first order transition, i.e. the phase boundary moves through the
sample providing the orienting effect on $\bf l$ and resulting in
stabilization of more anisotropic metastable 2D LIM state. Worthy
to mark that similar history dependent orientational effect on
$\bf l$ was observed in ABM phase of $^3$He in silica aerogel
\cite{Halp2}.

The ESP phases observed in \cite{we12} have been obtained in
N-aerogel sample which has not been squeezed in transverse
direction. In this case the 2D LIM state of the ESP1 phase should
be isotropic, but the above mentioned orienting effect should
remain resulting in the anisotropic 2D LIM state in the ESP2
phase.

\section{Orientation of orbital vector in N-aerogel}

The influence of aerogel deformation on a spatial distribution of
$\bf l$ in ABM-like phase of $^3$He is a complex problem and
depends on how $\bf l$-orienting centers are transformed during
deformation. Different types of aerogel have different microscopic
structures. This can result in a different response of the $\bf
l$-field to the deformation. For example, silica aerogels used in
\cite{Halp3,Halp4} orient $\bf l$ {\it along} the axis of
stretching and {\it normal} to the axis of squeezing. On the other
hand, N-aerogel (i.e. the infinitely stretched array of cylindric
strands) orients $\bf l$ normal to the strands, i.e. {\it normal}
to the stretching.

There are three theoretical models describing the influence of the
aerogel deformation on $\bf l$ \cite{Vol,Sur,Sauls}. The model
\cite{Vol} considers the aerogel as a system of randomly oriented
cylinders and seems to be the most consistent with N-aerogel. The
deformation changes an angular distribution of strands and orients
the $\bf l$-field. The model \cite{Vol} predicts that $\bf l$
tends to align along the axis of squeezing and normal to the axis
of stretching. N-aerogel corresponds to the infinite stretching
and the model predicts the 2D LIM state in the $\xi-\eta$ plane in
agreement with experiments \cite{we12} and with this work.

\begin{figure}[t]
\centerline{\includegraphics[width=2.6in]{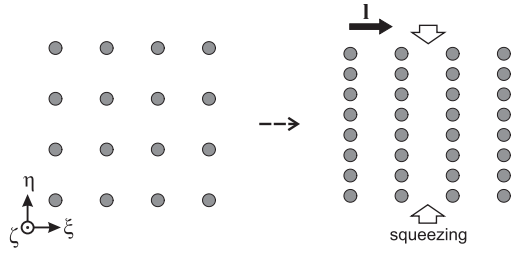}} \caption{Fig.7.
Squeezing of N-aerogel which strands form the 2D square lattice.}
\label{f7}
\end{figure}

As it was shown above the squeezing of the N-aerogel in the
$\hat\xi$-$\hat\eta$ plane results in preferable orientation of
$\bf l$ along the direction {\it normal} to the squeezing
direction. At first glance, this disagrees with \cite{Vol}.
However there is no contradiction here: the point is that the
squeezing in the $\hat\xi$-$\hat\eta$ plane does not change
orientations of the strands, i.e. the orienting effect in frames
of the model \cite{Vol} is absent. However, this type of
deformation changes spatial correlations of strands. If these
correlations are anisotropic in $\hat\xi$-$\hat\eta$ plane then
the orienting force will be along the direction normal to the
squeezing. This is illustrated by Fig.\ref{f7} where the result of
the squeezing of N-aerogel is shown for the simplest case of a 2D
square lattice of the strands in $\hat\xi$-$\hat\eta$ plane. It
can be seen that a strong squeezing results in the formation of
``wall-like'' structures. These ``walls'' should orient $\bf l$
normal to the surface, i.e. {\it normal} to the squeezing
direction. We think that this effect remains not only for the
cubic lattice but also for any locally anisotropic spatial
correlations. The similar phenomenon may cause the orienting
effect in a 3D lattice of ball-like $\bf l$-orienting centers if
their spatial correlations are locally anisotropic. We assume that
it may explain orienting effects observed in \cite{Halp3,Halp4}.

\section{Conclusions}
1. The observed NMR properties of ABM phase with polar distortion
of $^3$He in ``nematically ordered'' aerogel agree with the
developed theoretical model. This allows us to explain the
difference in NMR properties of the ESP phases: the 2D LIM state
of vector $\bf l$ in the ESP2 phase is more anisotropic than in
the ESP1 phase.

2. We have shown that the squeezing of N-aerogel along the
direction normal to the strands results in the anisotropic 2D LIM
state in the ESP phases so that the preferable orientation of $\bf
l$-field is normal to the squeezing. The explanation of this
effect is suggested. The anisotropy of the 2D LIM state decreases
on warming and may disappear below $T_{ca}$.

3. The order parameter orientational transition have been
observed. The transition occurs when the angle between $\bf H$ and
the axis of the anisotropy reaches the critical value.

This work was supported in part by RFBR (grant 13-02-00674) and by
RAS Program ``Quantum mesoscopic and disordered structures''. We
are grateful to I.A.\,Fomin and G.E.\,Volovik for useful comments.

\end{document}